\documentclass[
twocolumn,
showpacs,preprintnumbers,amsmath,amssymb,aps,pra,superscriptaddress,footinbib]{revtex4-1}

\usepackage{graphicx}
\usepackage{bm}
\usepackage{amsfonts}
\usepackage{amscd}
\usepackage{amsmath}    
\usepackage{enumerate}
\usepackage{subfigure}

\newcommand{\prlsection}[1]{\emph{#1}~---~}

\newcounter{mysectioncounter}

\usepackage{amssymb}
\usepackage{amsthm}

\newtheorem{theorem}{Theorem}

\newtheorem{corollary}{Corollary}

\newcommand{\bra}[1]{\mbox{$\left\langle #1 \right|$}}
\newcommand{\ket}[1]{\mbox{$\left| #1 \right\rangle$}}

\usepackage{color}
\definecolor{mycolor}{rgb}{1,0.5,0}
\definecolor{mycolor2}{rgb}{0,0.5,1}
\definecolor{mycolor3}{rgb}{0,0.8,.2}
\definecolor{mycolor4}{rgb}{1,0.2,.8}

\begin{document}

\title{
Universal Squash Model For Optical Communications Using
 Linear Optics And Threshold Detectors 
 }

\author{Chi-Hang Fred Fung}\email{chffung@hkucc.hku.hk}
\affiliation{Department of Physics and Center of Computational and Theoretical
 Physics, University of Hong Kong, Pokfulam Road, Hong Kong}
\author{H.~F. Chau}
\email{hfchau@hkusua.hku.hk}
\affiliation{Department of Physics and Center of Computational and Theoretical
 Physics, University of Hong Kong, Pokfulam Road, Hong Kong}
\author{Hoi-Kwong Lo}
\email{hklo@comm.utoronto.ca}
\affiliation{Center for Quantum Information and Quantum Control,
Department of Physics and Department of Electrical \& Computer Engineering,University of Toronto, Toronto, Ontario, M5S 3G4, Canada}

\date{\today}

\begin{abstract}

The transmission of 
photons
through open-air or an optical fiber is an important primitive in quantum information processing.
Theoretical description of such a transmission process often considers only a single photon as the information carrier and thus fails to accurately describe experimental optical implementations where any number of photons may enter a detector.
It is important
to bridge this big gap between experimental implementations and the theoretical description.
One powerful method
that emerges from recent efforts
to achieve this goal
is to consider a squash model that conceptually converts multi-photon states to single-photon states, thereby justifying the equivalence between theory and experiments.
However, up to now, only a limited number of protocols admit a squash model;
furthermore,
a no-go theorem has been proven which appears to rule out the existence of a universal squash model.
Here, we observe that
an apparently necessary condition demanded by all existing squash models
to preserve measurement statistics
is too stringent
a requirement 
for many protocols.
By chopping this requirement,
we show that
rather 
surprisingly, 
a universal squash model actually exists for a wide range of protocols 
including quantum key distribution protocols,
quantum state tomography,
the testing of Bell's inequalities,
and entanglement verification,
despite the standard no-go theorem.

\end{abstract}


\maketitle

\section{Introduction}

Quantum mechanics opens up new ways to process information.
Quantum information processing (QIP) allows tasks not possible in classical information processing, such as 
non-local correlations~\cite{Bell1964,Clauser1969}, and
unconditionally secure schemes for
cryptography~\cite{Bennett1984,Ekert1991,Gisin2002,*Lo2007_review}, randomness generation~\cite{Pironio2010}, and 
data hiding~\cite{PhysRevLett.86.5807,DiVincenzo2002}.
One of the greatest triumphs of QIP to date is 
quantum key distribution (QKD) (a.k.a. quantum cryptography), which 
allows two distant users to share a secret (as a classical bit string) by sending quantum states over a quantum channel.
Due to the ease of generation, transmission, and detection, photons are often used as information carrier in many quantum communication~\cite{Duan2001} tasks including
QKD (see, e.g., \cite{Brequet1994,muller:793,PhysRevLett.96.070502}),
teleportation (see, e.g., ~\cite{Bouwmeester1997,PhysRevLett.92.047904,Sherson2006,S.Olmschenk01232009}),
superdense coding (see, e.g., ~\cite{PhysRevLett.69.2881,PhysRevLett.76.4656,Barreiro2008}), and
quantum 
networks (see, e.g.,~\cite{Kimble2008,PhysRevLett.96.070504}).

In many quantum communication schemes (such as the most well-known QKD protocol -- the Bennett-Brassard-1984 protocol~\cite{Bennett1984} -- and quantum state tomography~\cite{Paris2004}),
the analyses often work on 
the assumption that
the quantum channel presents 
single-photon signals
to a receiver.
These signals
are subsequently measured with 
single-photon measurements.
However, in practice, experimental equipment fall short in guaranteeing such a pure 
single-photon environment.
This is because
practical photon sources occasionally emit more than one photon, and 
the detection setup for implementing the qubit measurement is usually composed of threshold detectors (such as standard InGaAs or silicon avalanche photo-diodes).
Threshold detectors only produce a click if the input signal contains one or more photons;
thus, they are incapable of revealing the number of photons entering the detection setup.
This immediately raises a key question:
does this mean that all single-photon-based quantum communication schemes  cannot run as expected from their original design and analyses?
For example, it is unclear whether a single-photon-based QKD protocol can still provide unconditional security when multi-photon signals are received from the quantum channel.
Also, in quantum state tomography, 
it is unclear whether we can ascribe a single-photon description to a state that we measure when such a state comes from a source that occasionally emits multi-photon signals.

The problem was initially motivated by QKD~\cite{Gottesman2004,Tsurumaru2008,Beaudry2008}%
, but was realized to be important in other QIP tasks such as entanglement verification~\cite{Beaudry2008,PhysRevA.81.052342}.
This is because many QIP tasks also rely on qubits as the basis of analysis but they do not carry the immediate security concern of QKD 
that
an intelligent eavesdropper may meticulously 
align her strategy with the practical detectors' behaviour.
Our goal is to bridge the idealization of qubit-based quantum communications and the physical realization where multi-photon signals may be emitted from the source and/or received from the quantum channel.
Indeed, the significance of this gap was demonstrated by Semenov and Vogel~\cite{Semenov2010} who showed that 
the mismatch between the theoretical single-photon consideration and the actual experimental reality with multi-photon signals might produce a fake violation of Bell's inequality and even quantum physics~\footnote{
The apparent violation is due to the 
discarding of double-click events and can be fixed by randomly assigning a bit value to them and keeping them.}.

While we address this QKD-motivated problem in the QKD context, our discussion applies equally to other contexts including quantum state tomography.
Indeed, later in the paper, we will show that quantum state tomography technique can be also be applied to a detection setup with threshold detectors.
Also, we note that there is a deep connection between the security of QKD and the testing of Bell's inequality which was first mentioned by Ekert~\cite{Ekert1991} in 1991 and was subsequently demonstrated through
the idea of ``self-testing'' for QKD
~\cite{Mayers1998,Mayers2004}, device-independent QKD based on Bell's inequality~\cite{PhysRevLett.98.230501,Pironio2009}, and state tomography based on Bell's inequality~\cite{PhysRevA.80.062327}.

QKD can be either prepare-and-measure or entanglement-based.
In the former, one party Alice prepares a quantum state and sends it to another party Bob who immediately measures it upon reception, and in the latter, an entanglement source generates a pair of entangled quantum states to be distributed to the two parties.
The multi-photon problem arises on the source side for prepare-and-measure QKD protocols because a phase-randomized weak coherent source is often used to simulate a single-photon source.
Since the multiple photons in an emitted signal 
are modulated to
carry the same information, extra copies of quantum information is available to Eve.
The multi-photon problem on the receiver side arises from the use of threshold detectors in both type of QKD schemes.

\begin{figure}[t]
\includegraphics
[width=1\columnwidth]
{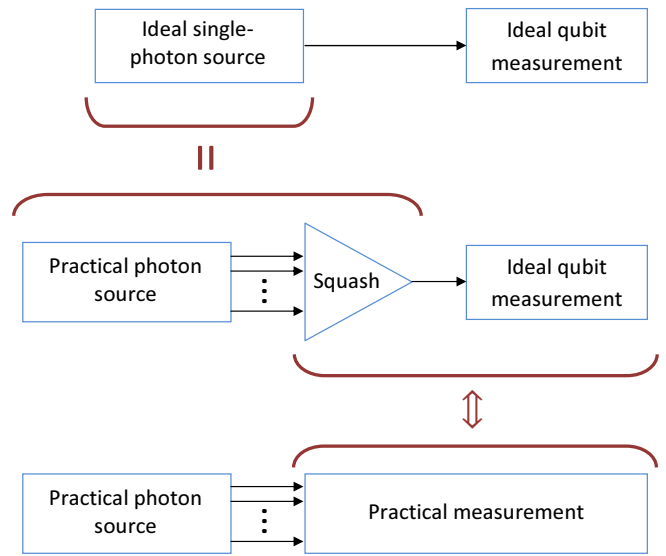}
\caption{\label{fig-squash-overview}
The squash operation is a valid quantum operation that maps a multi-photon input state 
to a single-photon output state.
The practical source together with the squash operation can be regarded as a single-photon source.
The practical measurement on the multi-photon signal consists of threshold detectors and is argued in this paper to be statistically related to the combination of a universal squash operation and an ideal qubit measurement.
Here, for simplicity, we use photon sources to also mean 
quantum channels.
}
\end{figure}

Existing efforts in setting up actual QKD experiments involving threshold
detectors and multiple-photon
sources as well as in proving the unconditional security of QKD schemes
using ideal apparatus~\cite{Mayers2001,Lo1999,Shor2000,Renner2005c,Renner2005}
will not be wasted if one can slightly modify the
post-processing procedure, the security proof, or the existing experimental
setup using currently available technologies.
Along this line of thought,
the multi-photon problem at the source (for prepare-and-measure schemes) was solved by 
Gottesman, Lo, L{\"{u}}tkenhaus, and Preskill (GLLP)~\cite{Gottesman2004} 
with great performance enhancement from using decoy 
states~\cite{Hwang2003,Lo2005,Wang2005a} (see also Ref.~\cite{Inamori2005}).

On the other hand, the threshold detection problem at the receiver (for both prepare-and-measure and entanglement-based schemes) can be solved by conceptually assuming a quantum operation before Bob that
maps Eve's multi-photon states to single-photon states.
GLLP
called this a squash operation~\cite{Gottesman2004} (see Fig.~\ref{fig-squash-overview}).
If an actual squashing device were concatenated to the multi-photon quantum channel, we would have an effective single-photon quantum channel emitting only single photons.
Such a physical squashing device would immediately make the receiving side of the experimental setup compatible with the single-photon-based analyses.
Although such a squashing device is impractical,
the squashing approach can still be used to justify the conceptual presence of a squashing device, which is not always possible.
Actually,
the squash operation was fully justified only in a few security proofs for the BB84~\cite{Tsurumaru2008,Beaudry2008} and the BBM92 protocols~\cite{Tsurumaru2008}.
These squash-operation-based security proofs are rather specific and do not provide a universal way to translate Eve's attack that outputs a multi-photon state to an attack that outputs a single-photon state.
In particular, a squash map is proved to not exist for the six-state QKD scheme with active basis selection~\cite{Beaudry2008} which is a scheme first introduced by Bennett \emph{et al.}~\cite{Bennett1984_sixstate} and later by Bru\ss~\cite{Bruss1998}.
In summary, previous works%
~\footnote{
We note that there are other ways to tackle the threshold detection problem.
In particular,
a separating approach was used to justify the use of threshold detectors specifically for the BBM92 protocol~\cite{Koashi2008_threshold} and
the BB84 protocol~\cite{Koashi2009comple,Koashi2006} without considering squash.
Also, Kato and Tamaki~\cite{Kato2010} proved the security of the six-state protocol with active basis selection and threshold detection by direct calculation of conditional entropy and application of Koashi's security proof based on complementarity scenario~\cite{Koashi2007_complementarity}.
}
show that a universal squash does not exist.
This is a highly disappointing result because it appears to mean that for each 
protocol, one has to prove its security by a different method because one cannot apply a universal squash operation.

Despite the previous no-go theorem, here we show that a universal squash actually exists.
The previous no-go theorem rests on a rather strong assumption that a squash map must be able to reproduce the precise measurement statistics (e.g., error rates).
Preserving the statistics is a rather stringent requirement.
The success of our approach lies in that we do not attempt to reproduce the exact statistics of the conceptual squash situation and we recognize that 
most quantum protocols do not need knowledge of exact statistics to function (bounds on statistics also suffice).
Indeed,
by relaxing this requirement and allowing a universal squash to produce only bounds on 
statistics,
we show that, rather surprisingly, a universal squash actually exists.
Consequently,
many security proofs for single-photon QKD protocols (with active or passive basis selection) including 
the six-state protocol~\cite{Bruss1998}, 
three-states protocols~\cite{Boileau2005,*Fung2006b},
the SARG04 protocol (with four or six states)~\cite{Scarani2004,*Tamaki2006}, 
the $N$-basis protocol~\cite{Koashi2005_discrete,*Shirokoff2007},
and protocols using decoy states~\cite{Hwang2003,Lo2005,Wang2005a},
multi-partite quantum cryptographic protocols~\cite{Chen2007_multi-partite},
reference-frame independent QKD~\cite{Laing2010},
two-way protocols~\cite{Gottesman2003} 
directly carry over to practical implementations with a weak-coherent-state source and threshold detectors.
This amounts to immense simplification.
We emphasize that, in addition to QKD, our work also applies to 
quantum state tomography,
the testing of Bell's inequalities,
and entanglement verification.
Later in this paper, we will discuss the application of our universal squash model to quantum state tomography 
and the testing of Bell's inequalities.
Here, we remark that while the protocol-specific squashing approach of Ref.~\cite{Beaudry2008} has also been applied to entanglement verification~\cite{PhysRevA.81.052342}, our approach has the advantage of being protocol independent (i.e., universal) and being applicable in many different contexts including quantum state tomography.

The organization of our paper is as follows.
We discuss our proof on universal squash model in Sec.~\ref{sec-universal-squash-model} followed by 
a discussion on the special treatment for the key bits in QKD in Sec.~\ref{sec-post-selection-QKD}.
We then apply our universal squash model to QKD protocols in Sec.~\ref{sec-examples},
quantum state tomography in Sec.~\ref{sec-tomography}, and the testing of Bell's inequality in Sec.~\ref{sec-bell-inequality}.
We conclude in Sec.~\ref{sec-conclusions}.

\newcommand{\myitem}[1]{\noindent$\bullet$
\parbox[t]{.95\columnwidth}{#1}
\vspace{4pt}
}

\section{Universal squash model}
\label{sec-universal-squash-model}

\begin{figure}[t]
\includegraphics
[width=.6\columnwidth]
{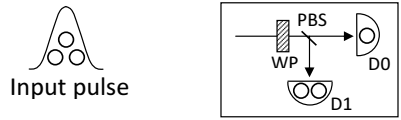}
\caption{\label{fig-basic-detection-system}
Detection system used by Bob for one basis, where
a set of waveplates (WP) select the basis and a polarizing beamsplitter (PBS) splits the signal into two arms for detection by two threshold detectors (D0 and D1).
Here, the incoming signal consists of three photons and one is (two are) collapsed in 
detector D0 (D1).
}
\end{figure}

We discuss our result assuming
the following settings:
\begin{itemize}


 \item
{
The incoming photons are restricted to a single optical spatio-temporal mode and information is encoding in polarization.
Note that there is no loss of generality in our discussion because phase encoding is mathematically equivalent to polarization encoding.
}

 \item
{
For simplicity, we assume that
Bob uses active basis selection for his measurements so that his detection system projects the incoming signal onto the eigenstates of only one basis.
(We extend it to the case of passive basis selection in Appendix~\ref{sec-passive-basis}.)
}

 \item
{
Bob's detection system 
consists of 
two
threshold detectors plus possibly other linear optical elements (a representative structure is shown in Fig.~\ref{fig-basic-detection-system}).
All photons in the same spatio-temporal mode entering each detector will be measured and collapsed individually.
In other words, the projection operators describing the measurement of each individual photon commute and are independent of each other.
Even though we focus on a two-detector receiver, for simplicity of discussion,
our proof works with any number of detectors where multiple clicks may occur. 
}

 \item
{
The threshold detectors have perfect efficiencies and no dark counts.
Therefore, all incoming photons are collapsed.
Inefficient detectors may be modeled as perfectly efficient detectors followed by a beamsplitter, which may be absorbed into the channel.
}

 \item
{
Quantum non-demolition (QND) measurements are implicitly assumed, without loss of generality, to be used by Bob to determine the input photon number throughout the proof.
The presence of the QND measurements is consistent with the threshold detector model and does not affect the functioning of it.
}
 \end{itemize}

\begin{figure*}
\includegraphics
{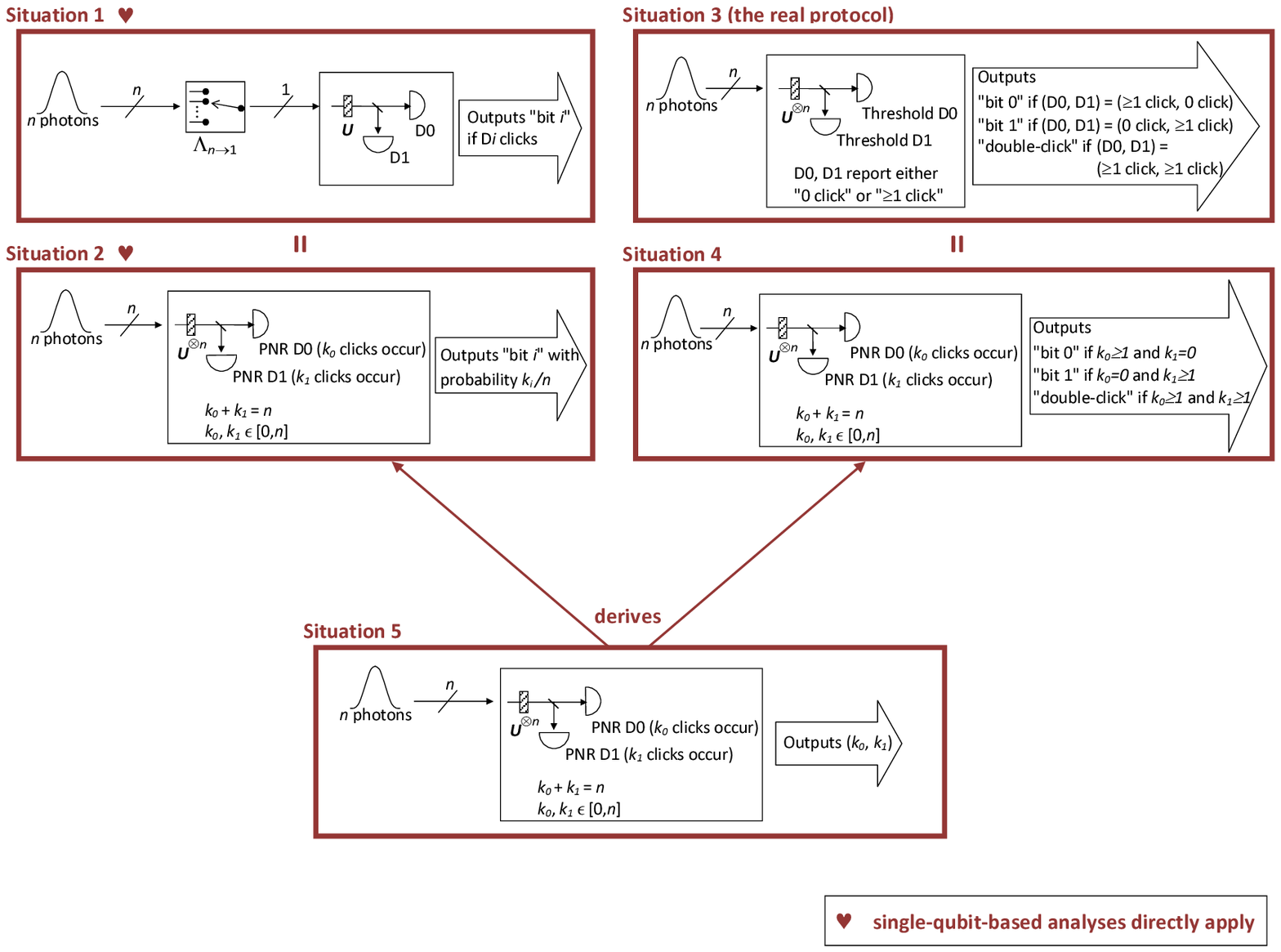}
\caption{\label{fig:situation_map_full}
Relationship map for the five situations used to prove our universal squash model.
The goal is to link the real situation (Situation 3) 
with the ideal situation (Situation 1) consisting of the universal squash operation.
\\
In Situation 1 (a virtual protocol),
an $n$-photon state enters a detection system comprising the squash operation $\Lambda_{n\rightarrow1}$, a set of waveplates (acting as unitary transform $U$), a polarizing beamsplitter, and two detectors.
The universal squash operation $\Lambda_{n\rightarrow1}$ maps $n$ photons to one. 
We do not specify whether the detectors are photon-number-resolving (PNR) or threshold detectors since after the squash operation, only one photon remains.
The output of the detection system is a bit value corresponding to the detector that has a click.
This is a single-qubit situation since we can regard the squash operation as part of the channel.
In Situation 2 (a virtual protocol), 
an $n$-photon state enters a detection system comprising a set of waveplates (acting as unitary transform $U^{\otimes n}$), a polarizing beamsplitter, and two 
PNR
detectors, followed by a classical post-processor.
The classical post-processor serves as the classical analog of the squash operation $\Lambda_{n\rightarrow1}$ and outputs a bit value according to probabilities given by the detectors' clicks.
More concretely, suppose that detectors D$0$ and D$1$ register $k_0$ and $k_1$ photons respectively.
Then, the classical post-processor in Situation 2 outputs event ``bit $i$'' ($i=0,1$) with  probability $k_i / (k_0 + k_1)$.
\\
We show in Theorem~\ref{thm-equivalent1} that the bit value outputs of Situations 1 and 2 have the same statistics for any $n$-photon input state and any unitary transform $U$ (see Supplementary Materials for proof).
\\
Situation 3 (the real protocol) is the similar to Situation 2 with the PNR detectors replaced by threshold detectors.
Situation 3 outputs event ``bit $i$'' if only detector 
D$i$
clicks ($i=0,1$) and event ``double-click'' if both detectors click.
Situation 4 is also similar to Situation 2
with a difference in the post-processing part.
Here, the post-processing only announces one of three events corresponding to a single-click for ``0'', a single-click for ``1'', and a double-click.
It is easy to see that the Situations 3 and 4 produce the same output statistics for the same input state.
In this paper,
we consider only those protocols for which Situations 3 and 4 are equivalent.
\\
Situation 5 is the mother protocol that derives Situations 2 and 4.
Note that the detection parts of Situations 2,4, and 5 are all the same; only their classical processing parts are different.
In fact, the classical processing parts of Situations 2 and 4 can be generated by that of Situation 5 which outputs the full information on the numbers of detection clicks.
}
\end{figure*}

Our proof can be illustrated pictorially as shown in Fig.~\ref{fig:situation_map_full}.
The essence
is to link the real situation (with threshold detectors and the possibility of double-click events) to an ideal situation (consisting of a universal squash operation), 
which are Situation 3 and Situation 1, respectively, in Fig.~\ref{fig:situation_map_full}.
This linking is established by regarding the two situations as statistically equivalent to special cases of classical post-processing for a detection setup with photon-number-resolving (PNR) detectors (Situations 2 and 4).
Both Situations 2 and 4 can be derived from Situation 5, which is a detection setup with PNR detectors that outputs the full information on the number of photons detected in each detector.
We discuss the elements of our proof as follows.

\subsection{State representation}
We write an $n$-photon pure state in tensor product form and then impose bosonic symmetry by symmetrizing the state~\footnote
{
We note that our formalism and result are fully consistent with the standard Hong-Ou-Mandel effect~\cite{PhysRevLett.59.2044} in quantum optics because we have imposed bosonic symmetry in our wave function.
}.
Similarly, an $n$-photon mixed state can be dealt with as a mixture of pure states.
Let $\rho$ denote the density matrix of an $n$-photon state.
A squash operation is a quantum operation that takes an $n$-photon state as input and produces a single-photon state as output.

\subsection{Universal squash operation}
We define our {\it universal squash operation} as the mapping from $\rho$ to $\rho_\text{qubit}$ where $\rho_\text{qubit}=\operatorname{Tr}(\rho)$ over any $n-1$ photons is the reduced density matrix of one photon.
It does not matter which $n-1$ photons we trace over and the same $\rho_\text{qubit}$ will result due to the bosonic symmetry.
We denote this mapping as $\Lambda_{n\rightarrow1}(\rho)=\rho_\text{qubit}$.
Note that $\Lambda_{n\rightarrow1}$ is a valid quantum operation.

\subsection{Equivalence of Situations 1 and 2}
\begin{theorem}
\label{thm-equivalent1}
{\rm
Situations 1 and 2 are equivalent and produce the same output statistics for any unitary transform 
and any $n$-qubit input state.
}
\end{theorem}
This result is non-trivial and its proof is discussed in Appendices~\ref{sec-proof-stat-eq} and \ref{app-generalized-measurements}.
The main point is that a universal squash operation (Situation 1) can be regarded as a special classical post-processing method for a detection system with PNR detectors (Situation 2).
Theorem~\ref{thm-equivalent1} justifies 
an effective single-qubit channel since the squash operation in Situation~1 can be regarded as part of the channel.
This means that it is valid to apply the result of any single-qubit-based analysis to Situation~2.

\subsection{Equivalence of Situations 3 and 4}
It is easy to see 
that the real situation with threshold detectors (Situation 3)  is equivalent to another special classical post-processing method for a detection system with PNR detectors (Situation 4).

\subsection{Relationship between Situations 2 and 4}
Situations 2 and 4 are related through Situation 5 (shown in
 Fig.~\ref{fig:situation_map_full}).
Since both Situations 2 and 4 are special cases of classical post-processing for a detection system with PNR detectors, they can both be derived from the same situation -- Situation 5.
Thus, the statistics of Situation 4 can be used to infer some statistics about Situation 2.
For QKD protocols, the statistics of interest is the error rate $e_b$ between Alice and Bob in basis $b$.
Thus, we aim at providing bounds on the error rates.
A single-click in Situation 4 immediately tells us that a 
definite
bit value would have been obtained in Situation 2 and 
this bit value is directly used for the evaluation of the error rates.
On the other hand, a double-click in Situation 4 does not tell us which bit value it corresponds to in Situation 2, 
and there is no definite bit value to be used for the error-rate evaluation.
To overcome this,
we recognize that we do not need to know the definite bit value since all we care are bounds on the error rates.
Our key idea is to bound the range of possible error rates by using the 
most pessimistic and optimistic values for double-click events.
Specifically, a double-click event contributes as an error bit for the calculation of the upper bound on the error rate and contributes as a correct bit for the lower bound.
Suppose that the number of test bits for basis $b$ is $N_b$, 
where
$N_b=N_b^{\text{s},\text{c}}+N_b^{\text{s},\text{e}}+N_b^{\text{d}}$. 
Here,
$N_b^{\text{s},\text{c}}$, $N_b^{\text{s},\text{e}}$, and $N_b^{\text{d}}$ are the correct single-click events, erroneous single-click events, and double-click events, respectively.
Then, the error rate of the test bits is bounded by
\begin{equation}
\label{eqn-error-rate-bounds}
e_b^\text{L}=
\frac{N_b^{\text{s},\text{e}}}{N_b} \leq e_b \leq \frac{N_b^{\text{s},\text{e}}+N_b^{\text{d}}}{N_b} 
=e_b^\text{U} .
\end{equation}

\begin{corollary}
\label{corollary-single-qubit}
{\rm
(Single-qubit description)
We regard the original quantum channel followed by the squash operation $\Lambda_{n\rightarrow1}$ in Situation 1 as the {\it effective single-qubit quantum channel}.
Thus, we can ascribe a single-qubit description to the actual received signals
 and the associated channel error statistics are bounded by Eq.~\eqref{eqn-error-rate-bounds}.
}
\end{corollary}
This allows us to apply any single-qubit-based security analysis to qubit-based QKD protocols whose qubit assumption is violated in practical implementation due to the reception of multi-photons.

Note that there are two such effective single-qubit quantum channels for entanglement-based QKD protocols (in which an entanglement source sends two signals one to Bob and one to Alice).

\section{Post-selection of key bits in QKD}\label{sec-post-selection-QKD}
In QKD protocols, 
one subset of the data obtained from the quantum channel (called the test bits) is devoted to  estimating the statistics of the channel and another subset (called the key bits) for 
key generation.
The key bits are eventually transformed into the final secret key through a series of classical operations and communications.

The bounds established above given in Eq.~\eqref{eqn-error-rate-bounds} apply to the channel output states irrespective of whether they are used as test bits or key bits.
Since we actually need to use the bit values of the key bits, we propose to discard all the double-click key bits for which we do not know the values.
This post-selection procedure requires us to extend the bounds of Eq.~\eqref{eqn-error-rate-bounds} when applied to the remaining key bits since we need to take into account the most pessimistic and optimistic error statistics of the discarded bits

More specifically,
if we had known the value of every key bit in Situation 2, we would have directly applied a qubit-based security proof to distill a final secret key.
However, we are in Situation 4 (or Situation 3, the real situation), and we do not know the key bit value for double-click events.
To solve this problem,
our strategy is to discard all double-click key bits and augment the error rate bounds of Eq.~\eqref{eqn-error-rate-bounds} for describing the remaining key bits.
Eventually, we will come up with new bounds for the error rates for the key after discarding:
\begin{equation}
\begin{aligned}
\label{eqn-error-rate-bounds-key1}
&e_{b}^\text{key,L} \leq e_{b}^\text{key} \leq e_{b}^\text{key,U}.
\end{aligned}
\end{equation}
To aid discussion, we designate one basis as the key-generating basis, denoted as basis $b^*$.
All key bits are detected in this basis and thus all single- or double-click events are classified according to this basis.
We first consider the $b$-basis error rate, where $b \neq b^*$.
Because the key bits discarded according to whether it is a double-click event in the $b^*$-basis may have any error rate in other bases,
the lower and upper bounds on the error rate ${e}_b^{\text{key}}$ in the $b$-basis for the remaining key bits are:
\begin{equation}
\label{eqn-error-rate-bounds-key-discard}
\begin{aligned}
e_b^\text{key,L}
&=
\frac{ e_b^\text{L} N_{b^*}^\text{key} - N_{b^*}^\text{key,d} } {N_{b^*}^\text{key,s}} \\
e_b^\text{key,U}
&=
\frac{ e_b^\text{U} N_{b^*}^\text{key} } { N_{b^*}^\text{key,s} }
, \text{ for }b\neq b^*
\end{aligned}
\end{equation}
where $ N_{b^*}^{\text{key},\text{s}} $ ($ N_{b^*}^{\text{key},\text{d}} $) is the number of single-click (double-click) events among all the $ N_{b^*}^\text{key} = N_{b^*}^{\text{key},\text{s}} + N_{b^*}^{\text{key},\text{d}} $ key bits measured in basis $b^*$.
Here, the lower (upper) bound 
comes from assuming that
the $b^*$-basis double-click events discarded correspond to erroneous (correct) bits in the $b$-basis.

The $b^*$-basis error rate of the key bits after discarding double clicks can be inferred from the $b^*$-basis error rate of the single-click test bits.
In the asymptotic case, these two error rates are the same, and are given by
excluding the double-click events in Eq.~\eqref{eqn-error-rate-bounds}:
\begin{eqnarray}
\label{eqn-error-rate-bounds-key2}
e_{b^*}^\text{key}=
\frac{N_{b^*}^{\text{s},\text{e}}}{N_{b^*}^{\text{s},\text{c}}+N_{b^*}^{\text{s},\text{e}}}.
\end{eqnarray}

Eqs.~\eqref{eqn-error-rate-bounds-key1}-\eqref{eqn-error-rate-bounds-key2} describe the key bits measured in the $b^*$-basis after discarding double-click events.
Classical post-processing steps derived from a qubit-based security proof can then be used to distill a final secret from the key bits according to these equations.
Also, we only need to consider physical states that satisfy the error rate constraints given in 
Eqs.~\eqref{eqn-error-rate-bounds-key1}-\eqref{eqn-error-rate-bounds-key2};
unphysical states need not be considered.

The key distillation steps are usually composed of two parts: error correction (EC) and privacy amplification (PA).
EC and PA are classical procedures 
for correcting errors between Alice's and Bob's initial keys and for removing Eve's information about their keys, respectively.
The codes/schemes and the associated parameters for EC and PA are determined from Eqs.~\eqref{eqn-error-rate-bounds-key1}-\eqref{eqn-error-rate-bounds-key2}
according to the security proof for the particular QKD protocol.

\subsection{QKD post-processing with error-rate ranges}
We emphasize that our method
involves EC and PA that operate on error rates given in ranges 
where the lower bound is not necessarily zero.
Note that most realistic EC codes are already designed to correct errors up to a certain error rate starting from zero.
Also, PA schemes based on phase error correction using random hashing or random codes~\cite{Mayers2001,Shor2000,Lo1999} and those based on information-theoretic proofs using universal hashing~\cite{Renner2005c,Renner2005} chosen according to the worst-case error rates automatically tolerate intermediate error rates.

Note that correlation between the different bases, which are reflected in the error rates and the structure of their respective bases (e.g., as in the six-state protocol), can be exploited by choosing the EC code and the PA scheme appropriately.
To compute the asymptotic key generation rate, it is sufficient to consider the worst-case physical state 
(see examples in Sec.~\ref{sec-examples}).

\section{QKD Examples}
\label{sec-examples}

\subsection{Six-state QKD protocol}
Considering the asymptotic case, suppose that 
Alice and Bob observe that their measurement results on the test bits for bases $X$, $Y$, and $Z$ all have the same erroneous single-click rate $\epsilon$, double-click rate $\delta$, and correct single-click rate $1-\delta-\epsilon$.
\footnote{
One such attack is the following.
For each qubit sent by Alice, Eve does nothing to the qubit with probability $1-p_1-p_2$ and processes the qubit with probability $p_i/3, i=1,2$ with the following copying operation in each of the bases $W=X, Y, Z$:
\begin{eqnarray*}
\alpha \ket{0_W} + \beta \ket{1_W} \rightarrow 
\alpha \ket{0_W}_B^{\otimes i} \ket{0_W}_E + \beta \ket{1_W}_B^{\otimes i} \ket{1_W}_E
\end{eqnarray*}
where $i=1,2$ output qubits are sent to Bob and one is kept by Eve.
Thus, when Eve launches the copying operation in $Z$ with $i=2$ and Bob detects in the basis $X$, he gets bit $0$, bit $1$, and a double click with probabilities $1/4$, $1/4$, and $1/2$, respectively.
The asymptotic error probabilities are bounded by
\begin{eqnarray*}
2\left(\frac{p_1}{6} + \frac{p_2}{12} \right) \leq e_X,e_Y,e_Z \leq 2\left(\frac{p_1}{6} + \frac{p_2}{12} + \frac{p_2}{6} \right) .
\end{eqnarray*}%
}
This induces the following bounds on the asymptotic error probabilities on the key bits (before discarding double-clicks) according to Eq.~\eqref{eqn-error-rate-bounds}:
\begin{eqnarray}
\epsilon \leq e_X,e_Y,e_Z \leq \epsilon+\delta .
\end{eqnarray}
The key bits are measured in the $Z$-basis and double-click key bits are discarded.
After discarding, the error rate in the $Z$-basis of the post-selected key bits is given by Eq.~\eqref{eqn-error-rate-bounds-key2}
due to direct inference from the test bits 
and the error rates in the other bases are bounded with Eq.~\eqref{eqn-error-rate-bounds-key-discard}:
\begin{equation}
\label{eqn-sixstate-constraints1}
\begin{aligned}
&{e}_Z^{\text{key}}=
\frac{\epsilon}{1-\delta} \\
\frac{\epsilon-\delta}{1-\delta}
&
\leq
{e}_X^{\text{key}},{e}_Y^{\text{key}}
\leq
\frac{\epsilon+\delta}{1-\delta} .
\end{aligned}
\end{equation}
The key generation rate 
for the six-state protocol with one-way error reconciliation is~(cf. Ref.~\cite{Lo2001_sixstate}):
\begin{eqnarray}
R_\text{six-state}&=&\min_{b_1,b_2,b_3} (1-\delta) \left[1-h\left(b_1,b_2,b_3,1-\sum_{i=1}^3 b_i\right)\right]
\nonumber \\
\label{eqn-example-sixstate-keyrate1}
\end{eqnarray}
subject to
${e}_Z^{\text{key}}=b_1+b_2$, ${e}_X^{\text{key}}=b_2+b_3$, ${e}_Y^{\text{key}}=b_1+b_3$,
and Eq.~\eqref{eqn-sixstate-constraints1}, 
where $h(b_1,b_2,\ldots)=-\sum_i b_i \log b_i$.
The minimum is achieved with, under some condition, 
(see Appendix~\ref{app-six-state-derivation})
\begin{eqnarray}
\label{eqn-sixstate-sol1a}
b_1&=&\frac{\epsilon}{1-\delta}-b_2 \\
b_3&=&\frac{\epsilon+\delta}{1-\delta}-b_2 \\
\label{eqn-sixstate-sol1c}
b_2&=&
\frac{\epsilon-2\delta}{2(1-\delta)} . 
\end{eqnarray}
Note that in the limit that $\delta$ goes to zero, $b_1 = b_2 =b_3$, and we recover the standard result in single-photon-based six-state QKD.
This shows that at small double-click rates, the six-state protocol  still gives a higher key rate than the BB84 protocol~\footnote{The single-photon six-state protocol achieves a tolerable bit error rate of $12.6\%$ with one-way error reconciliation whereas the single-photon BB84 protocol achieves $11.0\%$.}. 
In practice, the double-click rate with a weak-coherent-state source is usually pretty small, since the average source intensity is usually less than one photon per pulse and the fiber medium incurs signal loss.
Thus, applying our universal squash model to the six-state protocol in practice does not incur much loss due to the pessimistic error-rate estimation and the discarding of double-click key bits.
This shows the power of our universal squash model in security proofs.

\subsection{The BB84 protocol}
\begin{figure}[t]
\includegraphics
[width=\columnwidth]
{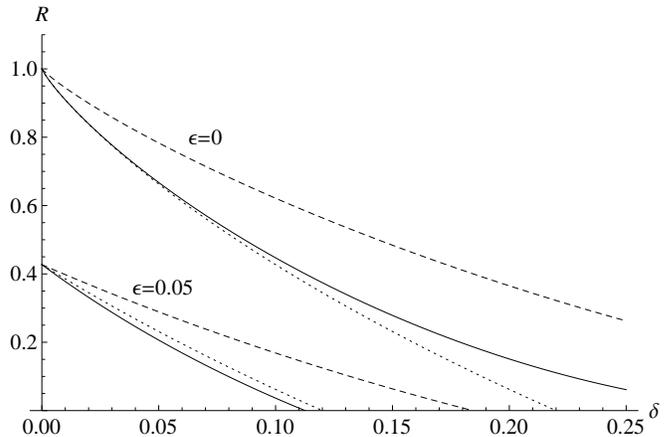}
\caption{\label{fig:bb84_comp}
Dependence of the key rate $R$ per detected signal on the double-click rate $\delta$.  The solid curves are for our universal squash model (Eq.~\eqref{eqn-BB84-rate1}) and the dotted curves and the dashed curves are for the BB84-specific squash model~\cite{Tsurumaru2008,Beaudry2008} (Eq.~\eqref{eqn-BB84-rate2} and Eq.~\eqref{eqn-BB84-rate3} respectively).
}
\end{figure}

Similar to the example on the six-state protocol, suppose that
Alice and Bob observe that their measurement results on the test bits for bases $X$, $Y$, and $Z$ all have the same erroneous single-click rate $\epsilon$, double-click rate $\delta$, and correct single-click rate $1-\delta-\epsilon$.
This induces the following bounds on the asymptotic error probabilities:
\begin{eqnarray}
\label{eqn-BB84-constraints1}
\epsilon \leq e_X,e_Z \leq \epsilon+\delta .
\end{eqnarray}
The key bits are measured in the $Z$-basis and double-click key bits are discarded.
After discarding, the error rate in the $Z$-basis of the post-selected key bits is $\epsilon$ due to direct inference from the test bits (cf. Eq.~\eqref{eqn-error-rate-bounds-key2}) and the error rate in the $X$-basis is bounded with Eq.~\eqref{eqn-error-rate-bounds-key-discard}:
\begin{equation}
\begin{aligned}
&
{e}_Z^{\text{key}}=
\frac{\epsilon}{1-\delta}
\\
\frac{\epsilon-\delta}{1-\delta}
&
\leq
{e}_X^{\text{key}} 
\leq
\frac{\epsilon+\delta}{1-\delta}
\end{aligned}
\end{equation}
The key generation rate with one-way reconciliation is thus~\cite{Lo1999,Shor2000,Renner2005,Kraus2005}
\begin{eqnarray}
R_\text{BB84}&=&(1-\delta)\left[1-
h_2\left(\frac{\epsilon}{1-\delta}\right)
-h_2\left(\frac{\epsilon+\delta}{1-\delta}\right)\right].
\nonumber\\
\label{eqn-BB84-rate1}
\end{eqnarray}
The squash model of Ref.~\cite{Tsurumaru2008,Beaudry2008} for the BB84 protocol preserve the measurement statistics when
the double-click events are randomly assigned a bit value.
Their key rate without discarding double-click events is
\begin{eqnarray}
\label{eqn-BB84-rate2}
R_\text{BB84}'&=&1-2h_2(\epsilon+\delta/2).
\end{eqnarray}
Figure~\ref{fig:bb84_comp} shows that our universal squash model given in Eq.~\eqref{eqn-BB84-rate1} has similar performance as
or even sometimes outperforms
the statistics-preserving squash model of Ref.~\cite{Tsurumaru2008,Beaudry2008} given in Eq.~\eqref{eqn-BB84-rate2}.
This is because we allow double-click key bits to be discarded in our model. 
For fair comparison, we also consider the key rate of the statistics-preserving squash model with discarding double-click key bits:
\begin{eqnarray}
R_\text{BB84}''&=&(1-\delta)\left[1-
h_2\left(\frac{\epsilon}{1-\delta}\right)
-h_2\left(\frac{\epsilon+\frac{\delta}{2}}{1-\delta}\right)\right]. \nonumber\\
\label{eqn-BB84-rate3}
\end{eqnarray}
Here, the first entropic term corresponds to bit error correction and the bit error rate is obtained with Eq.~\eqref{eqn-error-rate-bounds-key2}; the second entropic term corresponds to phase error correction and the phase error rate is upper bounded by assuming that all the double-click key bits discarded have no phase error and re-normalizing the phase error rate of the initial set of key bits $\epsilon+\frac{\delta}{2}$ by the size of the final set.
Figure~\ref{fig:bb84_comp} compares the two models, and it shows that
substantive difference exists between $R_\text{BB84}$ and $R_\text{BB84}''$ when the double-click rate is large.
However, in practice, the double-click rate is usually quite small.

\subsection{BB84 in realistic setting}

\begin{figure}
\subfigure[]{
\includegraphics[width=\columnwidth]{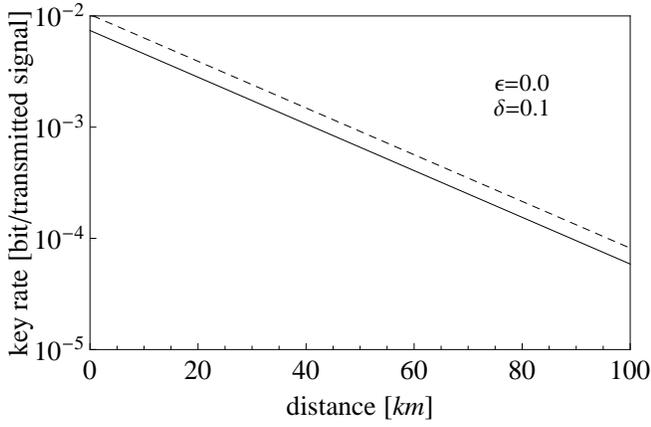}\label{fig:wcs_a}}
\\
\subfigure[]{
\includegraphics[width=\columnwidth]{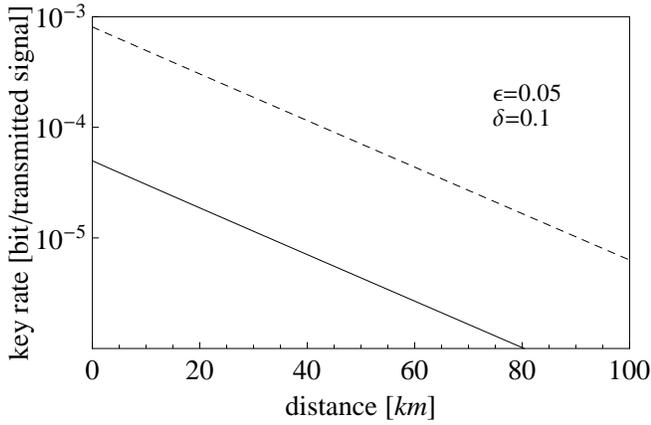}\label{fig:wcs_b}}
\\
\subfigure[]{
\includegraphics[width=\columnwidth]{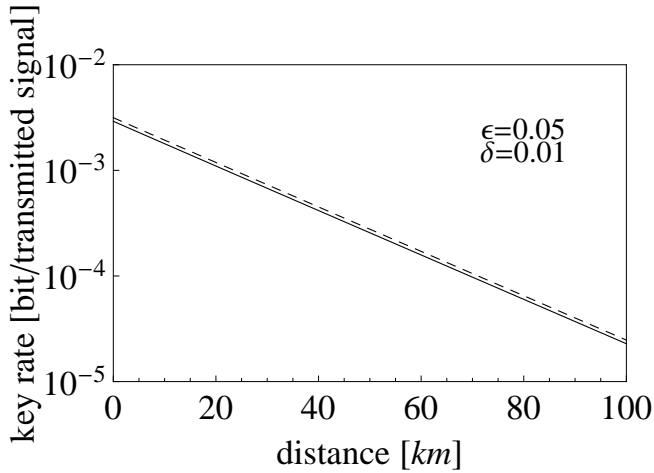}\label{fig:wcs_c}}
\caption{\label{fig:wcs}
Dependence of the key rate per transmitted signal on distance.  The solid curves are for our universal squash model (Eqs.~\eqref{eqn-decoy-universal-bound} and \eqref{eqn-decoy-key-rate}) and the dashed curves are for the BB84-specific squash model~\cite{Tsurumaru2008,Beaudry2008} (Eqs.~\eqref{eqn-decoy-random-assignment-bound} and \eqref{eqn-decoy-key-rate}).
Optimal $\mu$ is used at every distance.
Here, we used the experimental parameters for signals at wavelength $1550$ nm from the Gobby-Yuan-Shields experiments~\cite{GYS_04}: $\alpha=0.21$ dB/km and $\eta_\text{Bob}=4.5\%$ .
}
\end{figure}

We consider the performance of our universal squash model in the realistic setting with a weak coherent state source and the decoy-state method.
Suppose that Alice uses a phase-randomized weak-coherent-state source that emits signals of photon number $n$ with the Poisson distributed probability $p_{\mu,n}=\frac{e^{-\mu} \mu^n}{n!}$ where $\mu$ is the mean photon number.
Typically, $\mu$ is small and on the order of $1$ (e.g., $\mu = 0.5$).
The yield, $Y_n$, is the probability that Bob gets a detection (either a single-click event or a double-click event) given that Alice sent an $n$-photon signal.
For a state generated by the source with intensity $\mu$, the gain, defined as the probability that Bob has a detection and Alice sent an $n$-photon state, is
\begin{eqnarray}
Q_{\mu,n} &=& p_{\mu,n} Y_n.
\end{eqnarray}
The overall gain, single-click error rate, single-click correct rate, and double-click rate, are, respectively
\begin{eqnarray}
\label{eqn-decoy1}
Q_\mu &=& \sum_{n=0}^{\infty} p_{\mu,n} Y_n 
\\
Q_\mu F_\mu^\text{s,e} &=& \sum_{n=0}^{\infty} p_{\mu,n} Y_n F_n^\text{s,e}
\\
Q_\mu F_\mu^\text{s,c} &=& \sum_{n=0}^{\infty} p_{\mu,n} Y_n F_n^\text{s,c}
\\
\label{eqn-decoy2}
Q_\mu F_\mu^\text{d} &=& \sum_{n=0}^{\infty} p_{\mu,n} Y_n F_n^\text{d}
\end{eqnarray}
where
\begin{eqnarray}
F_n^\text{s,e}+F_n^\text{s,c}+F_n^\text{d}=1  \hspace{1cm} \forall n
\end{eqnarray}
and 
the three terms in the last equation are
the single-click error rate, single-click correct rate, and double-click rate
on $n$-photon states sent by Alice.

To estimate the channel parameters, we use the infinite decoy state protocol~\cite{Lo2005} in which we vary the source intensity $\mu$ continuously to generate infinitely many decoy states that form a system of linear equations from Eqs.~\eqref{eqn-decoy1}-\eqref{eqn-decoy2}.
We expect that a finite decoy state method will give similar results~\cite{Ma2005b,Wang2005a}.
The decoy states
are solely for parameter estimation,
and solving this system of equations gives $Y_n$, $F_n^\text{s,e}$, $F_n^\text{s,c}$, and $F_n^\text{d}$ for all $n$.
Since we now know the single-click error rate, single-click correct rate, and double-click rate for the single-photon states emitted by Alice, we can compute the worst-case single-photon error rates in our universal squash model in the same manner as in the previous analysis for single-photon sources.
According to Eq.~\eqref{eqn-error-rate-bounds},
they are 
bounded by
\begin{eqnarray}
F_1^\text{s,e} \leq e_{X,1}, e_{Z,1} \leq F_1^\text{s,e}+F_1^\text{d} .
\end{eqnarray}
The key bits are measured in the $Z$-basis and double-click key bits are discarded.
After discarding, 
the single-photon error rate in the $X$-basis is bounded as in  Eq.~\eqref{eqn-error-rate-bounds-key-discard} to be
\begin{eqnarray}
\label{eqn-decoy-universal-bound}
e_{X,1}^{\text{key}}
\leq
\frac
{F_1^\text{s,e}+F_1^\text{d}}
{1-F_1^\text{d}}  = e_{X,1}^{\text{key,U}}.
\end{eqnarray}
Since we correct all $Z$-basis errors for all states with any photon number,
the overall error rate in the $Z$-basis of the post-selected key bits is given by Eq.~\eqref{eqn-error-rate-bounds-key2} to be
\begin{eqnarray}
e_{Z,\bar{\mu}}^{\text{key}}=
\frac{ F_{\bar{\mu}}^\text{s,e} }
{F_{\bar{\mu}}^\text{s,e}+F_{\bar{\mu}}^\text{s,c}}.
\end{eqnarray}
Here, we use $\bar{\mu}$ to represent the intensity for the key-generating signal states.
The key rate, according to Gottesman-Lo-L{\"{u}}tkenhaus-Preskill~\cite{Gottesman2004}, is 
\begin{eqnarray}
R_\text{BB84,decoy}&=&-Q_{\bar{\mu}}(1-F_{\bar{\mu}}^\text{d}) h_2 \left( 
e_{Z,\bar{\mu}}^{\text{key}}
\right)
+ \nonumber \\
&&
Q_{\bar{\mu},1}(1-F_1^\text{d})
\left[ 1-
h_2\left(   
e_{X,1}^{\text{key,U}}
\right)
\right] . \phantom{xx}
\label{eqn-decoy-key-rate}
\end{eqnarray}

We assume the following simulation model to represent Eve's control over the channel parameters,
$Y_n$, $F_n^\text{s,e}$, $F_n^\text{s,c}$, and $F_n^\text{d}$.
Each photon has a certain transmission probability $\eta_\text{ch}=10^{-\frac{\alpha l}{10}}$ of not being lost in the fiber optic channel where $\alpha$ in dB/km is the loss coefficient of the fiber and $l$ in km is the length of the fiber.
Also, Bob's detectors have a certain detection efficiency $\eta_\text{Bob}$ for detecting an input photon.
Thus, the probability for a single photon to be detected by Bob is $\eta=\eta_\text{ch}\eta_\text{Bob}$.
The yield is
\begin{eqnarray}
Y_n &=& 1-(1-\eta)^n.
\end{eqnarray}
Since we do not assume the detectors to have dark counts, a single-photon signal emitted by Alice will at most produce one click on Bob's side if the signal goes through a passive channel.
Thus, for illustration purpose, we assume that Eve actively introduces multiple photons to Bob and her attack
induces $F_n^\text{s,e}=\epsilon$, $F_n^\text{d}=\delta$, and $F_n^\text{s,c}=1-\epsilon-\delta$ for all $n$.
Figure~\ref{fig:wcs} shows the key generation rates using this simulation model.
Results for our universal squash model and the 
statistics-preserving squash model of Ref.~\cite{Tsurumaru2008,Beaudry2008} specific to BB84 are shown.
The key rate for the latter case is given by Eq.~\eqref{eqn-decoy-key-rate} with
\begin{eqnarray}
\label{eqn-decoy-random-assignment-bound}
e_{X,1}^{\text{key,U}}
=
\frac
{F_1^\text{s,e}+F_1^\text{d}/2}
{1-F_1^\text{d}}  .
\end{eqnarray}
As shown in the figure, the performance degradation of our universal squash model is small when the single-click error rate $\epsilon$ is small or the double-click rate $\delta$ is small.
Note that there is no cutoff distance for both cases since we do not assume the detectors to have dark counts.

\section{Qubit state tomography}
\label{sec-tomography}
In many qubit state tomography experiments (e.g., \cite{PhysRevLett.83.3103,Shalm2009}), photons often are not generated from true single-photon sources.
When non-ideal sources and threshold detectors are used, 
without a squash model it is unclear whether one can talk about determining the qubit state since multi-photon signals may be emitted.
On the other hand, our universal squash approach allows quantum state tomography techniques to be rigorously applied even to detection setups with non-ideal sources and threshold detectors.
After a squash operation, a state is reduced to that of a qubit and one can determine tomographically
the state of the resulting qubit.
We emphasized that our universal squash model can also be applied to multi-qubit tomography~\cite{PhysRevA.64.052312}.
The overall argument for applying our universal squash model to tomography is similar to that for QKD, with a difference in the statistics of interest involved.
Here, as shown below, we are interested in the average measurement value instead of the error rate.
Standard qubit state tomography using the Stokes parameters~\cite{Stokes1852}
involves measuring the polarization of identical single-photon states $\rho$ in an ensemble using the three bases $X$, $Y$, and $Z$.
The Pauli matrices are $W=\ket{0_W}\bra{0_W}-\ket{1_W}\bra{1_W}$ where $W=X,Y,Z$, and
$\ket{b_X}=(\ket{0_Z}+(-1)^b \ket{1_Z})/\sqrt{2}$,
$\ket{b_Y}=(\ket{0_Z}+(-1)^b i \ket{1_Z})/\sqrt{2}$,
$b=0,1$.
The density matrix of the qubit $\rho$ can be tomographically determined to be
\begin{eqnarray}
\rho
&=&\frac{1}{2}I+\frac{1}{2} \sum_{W=X,Y,Z} \operatorname{Tr}(\rho W) W.
\label{eqn-tomography-full-state}
\end{eqnarray}

We now discuss how to bound the parameters $\operatorname{Tr}(\rho W)$ when non-ideal sources and threshold detectors are used.
Suppose that we measure basis $W$ with a threshold detection setup in Fig.~\ref{fig-basic-detection-system} for $N_{W}$ number of signals.
According to the measurement outcomes, we decompose the signals into three parts:
$N_{W}=N_{W}^{\text{s},+}+N_{W}^{\text{s},-}+N_{W}^{\text{d}}$ where
$N_{W}^{\text{s},\pm}$ and $N_{W}^{\text{d}}$ are the single-click events corresponding to the $\pm1$ outcome of the measurement $W$ and double-click events, respectively.
Then, 
$\operatorname{Tr}(\rho W)$ is bounded by
\begin{equation}
\begin{aligned}
\frac{N_{W}^{\text{s},+}-N_{W}^{\text{s},-}-N_{W}^{\text{d}}}{N_{W}} 
\leq 
\operatorname{Tr}(\rho W)
\\
\phantom{xx}\leq 
\frac{N_{W}^{\text{s},+}-N_{W}^{\text{s},-}+N_{W}^{\text{d}}}{N_{W}} 
.
\end{aligned}
\end{equation}
Using these bounds and $\operatorname{Tr}(\rho I)=1$, we can obtain a set of consistent qubit states using Eq.~\eqref{eqn-tomography-full-state}.
Essentially, our universal squash model allows one to ascribe a qubit description to the output states.

\section{Testing of Bell's inequality}
\label{sec-bell-inequality}

We can use our result to derive bounds on Bell's inequality~\cite{Bell1964} violation which may subsequently be used for qubit state tomography.
Here, the statistics of interest comes from the
Clauser-Horne-Shimony-Holt (CHSH) inequality~\cite{Clauser1969}
(perhaps the most famous Bell-type inequality~\cite{Bell1964})
which considers 
the statistics
$\chi=E[A_1 B_1]+E[A_1 B_2]+E[A_2 B_1]-E[A_2 B_2]$, where
$E[A_i B_j]=\bra{\psi} A_i \otimes B_j \ket{\psi}$ 
is the expectation value 
of the bipartite state $\ket{\psi}$
with $\{-1,+1\}$-valued observables $A_i$, $B_j$, $i,j=1,2$.
The maximum value of $\chi$ in quantum mechanics is $2\sqrt{2}$ which can be achieved by a maximally entangled state,
while
the maximum value of $\chi$ for states consistent with local hidden variable (LHV) models 
is $2$.
Thus, any experiment showing a violation of $\chi>2$ will rule out 
LHV
theories.
However, so far, no conclusive experimental violation%
~\cite{Freedman1972,Aspect1982,Weihs1998,Tittel1998,Rowe2001,Matsukevich2008,PhysRevLett.93.130409,PhysRevLett.91.110405,PhysRevA.49.3209,PhysRevLett.83.2872,PhysRevA.81.040101}
exists due to the difficulties in closing the locality, detection, and postselection loopholes.

It should be emphasized that for the purpose of ruling out LHV theories, any such experiments do not have to assume quantum mechanics to hold, let alone
a quantum channel that emits qubit signals.
In this case, when a double-click event occurs, one may apply any bit assignment scheme (such as a random bit assignment scheme) to this event without regard to the question of compatibility with any qubit squash model. 
On the other hand, our universal squash model becomes relevant when the CHSH violation is used in a quantum context.
We may consider the problem of characterizing a multi-photon pair source with respect to a perfectly entangled qubit pair source by checking the measurement statistics of the actual source.
This is in the spirit of 
the idea of self-testing quantum apparatus first proposed by
Mayers and Yao~\cite{Mayers1998,Mayers2004}. 
Along a similar line are 
device-independent QKD based on Bell's inequality~\cite{PhysRevLett.98.230501,Pironio2009} and state tomography based on Bell's inequality~\cite{PhysRevA.80.062327}.

To illustrate the applicability of our model to Bell's inequality testing,
let us consider quantum state tomography based on Bell's inequality~\cite{PhysRevA.80.062327}.
Bardyn {\it et al.}~\cite{PhysRevA.80.062327} have considered the problem of quantifying the closeness of an unknown entangled state emitted by a black box to ideal entangled states by simply testing the unknown state for its CHSH violation.
For entanglement sources that emit qubit pairs,
they showed that the fidelity $F$ for characterizing the closeness to  maximally entangled two-qubit states is bounded by 
\begin{eqnarray}
\label{eqn-relation-F-chi}
F\geq (1+\sqrt{[\chi_\text{obs}/2]^2-1})/2
\end{eqnarray}
where $\chi_\text{obs}$
is the observed CHSH violation.
Our result allows the application of Eq.~\eqref{eqn-relation-F-chi} even 
in situations where
practical entanglement sources 
that may emit pairs of multi-photon signals and threshold detectors are used.
In the end,
we may 
regard the practical source as
a two-qubit entangled source having certain fidelity to a maximally entangled state.

For the CHSH-inequality test, suppose that the number of bits for bases $A_i$ and $B_j$ is $N_{A_iB_j}$,
which can be decomposed as
$N_{A_iB_j}=N_{A_iB_j}^{\text{s},+}+N_{A_iB_j}^{\text{s},-}+N_{A_iB_j}^{\text{d}}$ where
$N_{A_iB_j}^{\text{s},\pm}$ and $N_{A_iB_j}^{\text{d}}$ are the single-click events corresponding to the $\pm1$ outcome of the measurement $A_i \otimes B_j$ and double-click events, respectively.
Then, $E[A_iB_j]$ is bounded by
\begin{equation}
\label{eqn-CHSH-bounds}
\begin{aligned}
E_{ij}^\text{L}=
\frac{N_{A_iB_j}^{\text{s},+}-N_{A_iB_j}^{\text{s},-}-N_{A_iB_j}^{\text{d}}}{N_{A_iB_j}} 
\leq E[A_iB_j] \\
\phantom{xxxxxxx}\leq 
\frac{N_{A_iB_j}^{\text{s},+}-N_{A_iB_j}^{\text{s},-}+N_{A_iB_j}^{\text{d}}}{N_{A_iB_j}} 
=E_{ij}^\text{U} ,
\end{aligned}
\end{equation}
and the CHSH violation $\chi$ can be bounded by combining the corresponding bounds for the various $E[A_iB_j]$'s:
\begin{equation}
\label{eqn-CHSH-bounds2}
\begin{aligned}
E_{11}^\text{L}+E_{12}^\text{L}+E_{21}^\text{L}-E_{22}^\text{U} \leq
\chi \phantom{xxx} \\
\leq
E_{11}^\text{U}+E_{12}^\text{U}+E_{21}^\text{U}-E_{22}^\text{L} .
\end{aligned}
\end{equation}

\begin{corollary}
{\rm
(CHSH-based source estimation)
We regard the original entangled state 
processed by the squash operation $\Lambda_{n\rightarrow1}$ on each side of the state in Situation 1 as the {\it effective two-qubit state}.
We bound the CHSH violation of the effective two-qubit state using Eqs.~\eqref{eqn-CHSH-bounds}-\eqref{eqn-CHSH-bounds2} and infer its fidelity with Eq.~\eqref{eqn-relation-F-chi}.
}
\end{corollary}

\section{Conclusions}
\label{sec-conclusions}
The use of threshold detectors has been a major obstacle in bridging the practical experiments on quantum protocols 
and their theoretical qubit-based analyses.
In this paper, we provide a universal solution to this for a wide range of protocols including single-qubit-based schemes for QKD, quantum state tomography, and entanglement verification.
This allows the translation of existing analyses that assume single-photon inputs to ones that can handle multiple-photon inputs detected with threshold detectors.
For future work, it will be interesting to explore the applicability of our universal squash model in other contexts such as 
quantum metrology~\cite{PhysRevLett.96.010401} and linear optics quantum computation~\cite{RevModPhys.79.135}.

\prlsection{Acknowledgments}
We thank enlightening discussions with Norbert L{\"{u}}tkenhaus and Kiyoshi Tamaki.
This work is supported by
RGC grants No.~HKU 701007P and 700709P of the HKSAR Government, the CRC program, CIFAR, NSERC, and QuantumWorks.

\appendix

\section{Proof of Theorem~\ref{thm-equivalent1}}

\subsection{Statistical Equivalence of Situations 1 and 2 for projective measurements\label{sec-proof-stat-eq}}
We show that the bit value outputs of Situations 1 and 2 have the same statistics for any $n$-photon input state and any unitary transform.
This result is non-trivial since it means that the quantum squash operation can be perfectly substituted by a classical operation.
We show this by proving that (i) the squash operation commutes with the unitary transform and (ii) directly verifying that the statistics of a single-photon detection after squash is the same as the statistics of a multi-photon detection followed by our classical post-processor.

For (i), 
it follows from the fact that the original state
$\rho$ lives in a tensor product space of $n$ qubits and
standard linear optics transformations act on each photon separately.
Thus, when the transformation is characterized by a unitary transform $U$ on one qubit, the transformation on the $n$-qubit state is $\rho':=U^{\otimes n} \rho (U^\dagger)^{\otimes n}$.
It can easily be checked that 
$\Lambda_{n\rightarrow1}(
\rho'
) = U \Lambda_{n\rightarrow1}(\rho) U^\dagger$.
Here, $U^\dagger$ denotes the adjoint of $U$.

For (ii),
we consider whether the statistics of a single-photon detection after squashing $\rho'$ is the same as the statistics of a PNR detection on $\rho'$ followed by our classical post-processor.
We can verify this by comparing the probabilities of producing bits ``0'' and ``1'' in the two cases.
In some sense this means that squash commutes with the final detection.
Due to the bosonic symmetry, the state $\rho'$ is
symmetric on exchange of the photons (e.g., $\bra{001}\rho'\ket{001}=\bra{010}\rho'\ket{010}=\bra{100}\rho'\ket{100}$ for a $3$-photon state $\rho'$).
In light of this symmetry, we denote the probability of $\rho'$ collapsing to $\ket{\mathbf{x}}$ in a PNR detection as $\lambda_{n-k,k}:=\bra{\mathbf{x}}\rho'\ket{\mathbf{x}}$ where $\mathbf{x}$ is an $n$-bit string containing $n-k$ ``0''s and $k$ ``1''s.
For example, $\bra{001}\rho'\ket{001}=\bra{010}\rho'\ket{010}=\bra{100}\rho'\ket{100}=\lambda_{2,1}$.
Normalization gives $\sum_{k=0}^n \binom{n}{k} \lambda_{n-k,k}=1$.

We now consider the outcome probabilities 
in Situation 2
where 
the detectors are PNR and their results are processed by a classical post-processor whose behaviour is defined in 
Fig.~\ref{fig:situation_map_full}.
In general, for an $n$-photon state, 
\begin{eqnarray}
p_0^{CP}&=&\sum_{k=0}^n \binom{n}{k} \frac{n-k}{n} \lambda_{n-k,k} \\ p_1^{CP}&=&\sum_{k=0}^n \binom{n}{k} \frac{k}{n} \lambda_{n-k,k}.
\end{eqnarray}

Next, we consider the outcome probabilities 
in Situation 1 where a squash is
followed by a single-photon detection.
The probability of getting $0$ is
\begin{eqnarray}
p_0^{SQ}
&=&
\bra{0}\Lambda_{n\rightarrow1}(\rho')\ket{0}.
\end{eqnarray}
Expanding the partial trace of the squash operation with the assumption of tracing over qubits $2$ to $n$, we get
\begin{eqnarray}
p_0^{SQ}
&=&
\sum_{i_2,\ldots,i_n=0,1}
\bra{0 i_2 \cdots i_n}\rho'\ket{0 i_2 \cdots i_n}\\
&=&
\sum_{k=0}^{n-1} \binom{n-1}{k} \lambda_{n-k,k}.
\end{eqnarray}
Similarly, the probability of getting $1$ is
\begin{eqnarray}
p_1^{SQ}&=&\sum_{k=0}^{n-1} \binom{n-1}{k} \lambda_{k,n-k}.
\end{eqnarray}
It can be easily checked that $p_0^{CP}=p_0^{SQ}$ and $p_1^{CP}=p_1^{SQ}$ for all $n$.
This means that the squash operation commutes with the final detection.
Since (i) works for any arbitrary basis, 
this proves Theorem~\ref{thm-equivalent1}.
Note that here we only focus on projective measurements.
However, Theorem~\ref{thm-equivalent1} also holds for generalized measurements (see Subsection~\ref{app-generalized-measurements} next).

Had we used PNR detectors in practice (i.e., Situation~2), we would apply Theorem~\ref{thm-equivalent1} to argue that we had a quantum channel that emits a single qubit to be detected since the squash operation in Situation~1 in connection with the original multi-qubit channel can be regarded as an effective single-qubit channel.
This means that it is valid to apply the result of any single-qubit-based analysis.
However, when we use threshold detectors instead of PNR detectors, we have to make additional arguments, through Situations 3, 4, and 5, to justify the use of single-qubit-based analyses.

\subsection{Statistical equivalence of Situations 1 and 2 for generalized measurements\label{app-generalized-measurements}}

We show that Theorem~\ref{thm-equivalent1} also holds for generalized measurements by extending the proof of the previous Subsection~\ref{sec-proof-stat-eq}.
Suppose that the detection setup in Situations 1 and 2 is a POVM $\{ M_i, i=1,\ldots,m \}$ on a qubit instead of a simple projection onto $\{ \ket{0}\bra{0},\ket{1}\bra{1} \}$.
The $i$th PNR detector in Situation 2 indicates the number of clicks $n_i$ for the POVM element $M_i$.
Due to the bosonic symmetry of the $n$-photon state $\rho'$~\footnote{we consistently use the same notation $\rho'$ here to denote the state after a unitary transformation corresponding to active basis selection}, we denote the probability of collapsing it to $(x_1,\ldots,x_n)$ 
in the PNR detection in Situation 2 as $\lambda_{n_1,\ldots,n_m} := \operatorname{Tr}( M_{x_1}\otimes \cdots \otimes M_{x_n} \rho')$ where 
$x_1,\ldots,x_n$ each contains a POVM element index $[1,m]$ and there are $n_i$ number of them with index $i$.
For example, $\operatorname{Tr}( M_1\otimes M_1 \otimes M_2 \otimes M_3 \rho') = \operatorname{Tr}( M_3\otimes M_1 \otimes M_2 \otimes M_1 \rho') = \lambda_{2,1,1}$ for a $3$-element POVM ($m=3$) and a $4$-photon state ($n=4$).
Normalization gives 
\begin{eqnarray}
\sum_{
\substack{
n_1+\cdots+n_m=n\\
n_1,\ldots,n_m\geq0
}
} \binom{n}{n_1,\ldots,n_m} \lambda_{n_1,\ldots,n_m} =1.
\end{eqnarray}

We define the classical post-processor in Situation 2 to output outcome $i \in [1,\ldots,m]$ with probability 
$n_i/n$, when there are $n_i$ photons detected in the $i$th PNR detector.
Thus, the probability of getting outcome $i$ in Situation 2 is 
\begin{eqnarray}
\label{eqn-generalized-prob-SQ}
p_i^{CP}=
\sum_{
\substack{
n_1+\cdots+n_m=n\\
n_1,\ldots,n_m\geq0
}
} \binom{n}{n_1,\ldots,n_m} \frac{n_i}{n} \lambda_{n_1,\ldots,n_m}.
\end{eqnarray}

We now turn to the probability of getting outcome $i$ in Situation 1:
\begin{eqnarray}
p_i^{SQ} &=&
\operatorname{Tr}( M_i \Lambda_{n\rightarrow1}(\rho') ) .
\end{eqnarray}
We arbitrarily choose to trace over photons $2$ to $n$:
\begin{equation}
\begin{aligned}
\Lambda_{n\rightarrow1}(\rho')
=&
\operatorname{Tr}_{2,\ldots,n} ( \rho') \\
=&
\operatorname{Tr}_{2,\ldots,n} \left( 
\sum_{
\substack{
x_2,\ldots,x_n \\ \in [1,m]}}
M_{x_2}\otimes \cdots \otimes M_{x_n} \rho' \right) .
\end{aligned}
\end{equation}
Thus, we have
\begin{eqnarray}
\!\!\!\!\!\!
p_i^{SQ} &=&
\sum_{
\substack{
x_2,\ldots,x_n \\ \in [1,m]}}
\operatorname{Tr}( M_i \otimes M_{x_2}\otimes \cdots \otimes M_{x_n} \rho')
\\
&=&
\sum_{
\substack{
n_1+\cdots+n_m=n-1\\
n_1,\ldots,n_m\geq0
}}
\!\!
\binom{n-1}{n_1,\ldots,n_m} \lambda_{n_1,\ldots,n_i+1,\ldots,n_m}
\nonumber
\end{eqnarray}
which can easily be verified to be equal to Eq.~\eqref{eqn-generalized-prob-SQ}.
That is $p_i^{CP}=p_i^{SQ}$ for $i=1,\ldots,m$.

\section{Derivation for the six-state QKD protocol\label{app-six-state-derivation}}

The constraints for the six-state protocol are
\begin{eqnarray}
{e}_Z^{\text{key}}&=&b_1+b_2 , \\
{e}_X^{\text{key}}&=&b_2+b_3 , \\
{e}_Y^{\text{key}}&=&b_1+b_3 .
\end{eqnarray}
We first find $b_2$ as follows:
\begin{eqnarray}
b_2 &=& 
\frac
{{e}_Z^{\text{key}}+{e}_X^{\text{key}}-{e}_Y^{\text{key}}}
{2}
\end{eqnarray}
which, using Eq.~\eqref{eqn-sixstate-constraints1}, implies
\begin{eqnarray}
\label{eqn-sixstate-b2-bounds}
\frac{\epsilon-2\delta}{2(1-\delta)}
\leq
b_2
\leq
\frac{\epsilon+2\delta}{2(1-\delta)}.
\end{eqnarray}
The entropy term in the key rate expression in Eq.~\eqref{eqn-example-sixstate-keyrate1} can be broken down into a sum of a bit-error-correction term and a phase-error-correction term:
\begin{eqnarray}
h\left(b_0,b_1,b_2,b_3\right)
=
H(Z)+H(X|Z)
\end{eqnarray}
where
\begin{eqnarray}
H(Z)&=&h_2({e}_Z^{\text{key}})
\\
H(X|Z)&=&(1-{e}_Z^{\text{key}}) h_2\left(\frac{b_3}{1-{e}_Z^{\text{key}}} \right)
+ \nonumber
\\
&&{e}_Z^{\text{key}} h_2\left(\frac{b_2}{{e}_Z^{\text{key}}} \right)
\label{eqn-sixstate-H(X|Z)}
\\
b_0&=&1-\sum_{i=1}^3 b_i
\\
h_2(x)&=&-x \log x -(1-x) \log (1-x) .
\end{eqnarray}
We can substitute $b_3={e}_X^{\text{key}}-b_2$ in Eq.~\eqref{eqn-sixstate-H(X|Z)} and maximize $H(X|Z)$ over $b_2$ given its range in Eq.~\eqref{eqn-sixstate-b2-bounds} for fixed ${e}_Z^{\text{key}}$ and ${e}_X^{\text{key}}$.
Without the constraint of Eq.~\eqref{eqn-sixstate-b2-bounds} (i.e., the BB84 case), the maximizing value is $b_2={e}_Z^{\text{key}} {e}_X^{\text{key}}$.
For simplicity, we assume that $2\delta \ll \epsilon \ll 1/2$ so that the lower bound of $b_2$ in Eq.~\eqref{eqn-sixstate-b2-bounds} is greater than the upper bound of ${e}_Z^{\text{key}} {e}_X^{\text{key}}$.
Since $H(X|Z)$ is a concave function in $b_2$, this lower bound of $b_2$ is the maximizing value when taking into account the constraint of Eq.~\eqref{eqn-sixstate-b2-bounds}.
Next, we choose the largest possible value for ${e}_X^{\text{key}}$ in order to maximize $H(X|Z)$.
Thus, we arrive at Eqs.~\eqref{eqn-sixstate-sol1a}-\eqref{eqn-sixstate-sol1c}.

\section{Passive basis selection\label{sec-passive-basis}}

\begin{figure}[t]
\includegraphics
[width=.75\columnwidth]
{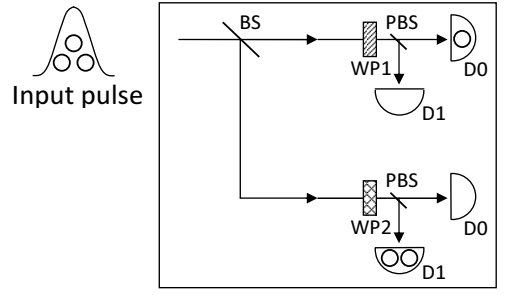}
\caption{\label{fig-passive-basic-detection-system}
Detection system used by Bob with passive basis selection among two bases, where
a beamsplitter (BS) splits the incoming signal into two paths corresponding to the two bases,
two sets of waveplates (WP1 and WP2) each select one of the bases, and a polarizing beamsplitter (PBS) splits the signal into two arms for detection by two threshold detectors (D0 and D1).
Here, the incoming signal consists of three photons and one is (two are) collapsed in 
detector D0 of the first basis (detector D1 of the second basis).
}
\end{figure}

We only consider uniform basis detection here.
This means that for a $B$-basis scheme, the initial received light path is split equally into $B$ paths.
A passive detection setup consists of orthogonal projections in multiple bases (see Fig.~\ref{fig-passive-basic-detection-system}).
We can regard that there is only one POVM measurement having many elements.
In the case of passive BB84, the POVM is $\{ \frac{1}{2}\ket{0_z}\bra{0_z}, \frac{1}{2}\ket{1_z}\bra{1_z}, \frac{1}{2}\ket{0_x}\bra{0_x}, \frac{1}{2}\ket{1_x}\bra{1_x} \}$.
Due to Appendix~\ref{app-generalized-measurements}, the statistics of Situations 1 and 2 are equivalent even in the passive-basis-selection case.
This establishes one link in the overall argument depicted in Fig.~\ref{fig:situation_map_full}.
The next step is to establish the other link --- the relation between the statistics that are actually observed in Situation 3 (or equivalently Situation 4) and the statistics of the ideal Situation 2.
Unlike the active-basis-selection case, the statistics of Situation 2 here have two parts: one for the error rates of the bit values, and one for the statistics of the basis values.
Here, we separately consider the two.

We argue that the statistics of basis values in Situation 4 is the same as that of Situation 1 (or equivalently Situation 2), when we adopt a particular basis selection rule for multi-basis events.
Thus, once a basis is selected with this rule, it only remains to 
consider the relation for the bit-value statistics.
Fortunately, this second part is the same as that of the active-basis-selection case and we can reuse the previous argument to estimate the error rates with the test bits and discard the double-click key bits.

Now we prove that the basis statistics of Situation 4 with the basis selection rule is the same as the basis statistics of Situation 1.
Obviously, in Situation 1, the single photon emitted from the squash operation collapses in each of the $B$ bases with probability $1/B$.
In Situation 4, we use the following basis selection rule:
\begin{itemize}
\item When only one basis has detection, we choose this basis.
\item When $b \leq B$ bases have detection, we choose one among these $b$ bases with uniform probability $1/b$.
\end{itemize}
The main point is to show
that a basis $A$ chosen according to this rule occurs with probability $1/B$, i.e., same as Situation 1.
This probability for an $n$-photon input state is as follows:
\begin{eqnarray}
P(A):=&&\sum_{b=1}^{B} \frac{1}{b} \binom{B-1}{b-1} \operatorname{Pr}\{  \text{detection in $b$ bases } \nonumber \\
&&\phantom{xxxxxxxxxxxxxx}\text{including basis $A$}\}
\phantom{xx}
\label{eqn-passive-prob-rule1}
\end{eqnarray}
where
\begin{eqnarray}
\!\!
P_b
&:=&\operatorname{Pr}\{  \text{detection in $b$ bases including basis $A$} \} \nonumber \\
&=&
\left(\frac{b}{B}\right)^n - \sum_{c=1}^{b-1} \binom{b}{b-c} P_{b-c}
\label{eqn-passive-prob-rule2}
\end{eqnarray}
Here, 
$P_b$ is defined in terms of a fixed set of $b$ bases of which basis $A$ is an element.
The first term in Eq.~\eqref{eqn-passive-prob-rule2} represents the probability that each of the $n$ photons collapses in any one basis of this set.
Since this term also includes the events that the $n$ photons collapse in less than $b$ bases, we exclude these events in the second term.

Rewriting Eq.~\eqref{eqn-passive-prob-rule1} as
\begin{eqnarray}
P(A)=&&\sum_{b=1}^{B-1} \frac{1}{b} \binom{B-1}{b-1} P_b + \frac{1}{B}P_B
\end{eqnarray}
and substituting the expression for $P_B$ from Eq.~\eqref{eqn-passive-prob-rule2}, we get
\begin{eqnarray}
P(A)=\frac{1}{B}.
\end{eqnarray}

This means that by adopting this basis selection rule, 
the basis statistics of Situation 1 is preserved.
The approach to deal with the bit-value statistics is the same as the active-basis-selection case and
we can reuse that result 
to handle the double-click events and to estimate the error rates.

\bibliographystyle{apsrev4-1}

\bibliography{paperdb}

\end{document}